\documentclass[aps,prd,12pt,showpacs,nofootinbib,tightenlines]{revtex4}
\usepackage{amsmath}
\usepackage{amssymb}
\usepackage{slashed}
\usepackage{epsfig}
\usepackage{graphicx}
\usepackage{enumerate}
\usepackage{multirow}
\usepackage[dvips,usenames]{color}
\usepackage{bm}% bold math
\begin{document}
%%%%%%%%%%%%%%%%%%%%%%%%%%%%%%%%%%%%%%%%%%%%%
%%%%%%%%%%%%%%%%%%%%%%%%%%%%%%%%%%%%%%%%%%%%%

\newcommand{\beq}{\begin{eqnarray}}
\newcommand{\eeq}{\end{eqnarray}}
\newcommand{\non}{\nonumber\\ }

\newcommand{\mb}{m_B }
\newcommand{\mtp}{m_{t'}}
\newcommand{\mw}{m_W }
\newcommand{\im}{{\rm Im} }

%%---------------------------------------------------------

\def \cpc{ {\bf Chin. Phys. C} }
\def \csb{ {\bf Chin. Sci. Bull.} }
\def \ctp{ {\bf Commun. Theor. Phys. } }
\def \ijmpa{ {\bf Int. J. Mod. Phys. A}  }
\def \copc{ {\bf Comput. Phys. Commum. } }
\def \epjc{{\bf Eur. Phys. J. C} }
\def \jpg{ {\bf J. Phys. G} }
\def \npb{ {\bf Nucl. Phys. B} }
\def \plb{ {\bf Phys. Lett. B} }
\def \pr{  {\bf Phys. Rep.} }
\def \rp{  {\bf Rev. Phys.} }
\def \prd{ {\bf Phys. Rev. D} }
\def \prl{ {\bf Phys. Rev. Lett.}  }
\def \ptp{ {\bf Prog. Theor. Phys. }  }
\def \rmp{ {\bf Rev. Mod. Phys. }  }
\def \zpc{ {\bf Z. Phys. C}  }
\def \jhep{ {\bf JHEP}  }
\def \epl{ {\bf EPL} }
\def \appb{ {\bf Acta Phys. Polon. B} }
%%%%%%%%%%%%%%%%%%%%%%%%%%%%%%%%%%%%%%%%%%%%%%%%%%%%
%%
\title{Searches for top-Higgs FCNC couplings via the $Whj$ signal with $h\to \gamma\gamma$ at the LHC}
\author{Yao-Bei Liu$^{1}$\footnote{E-mail: liuyaobei@sina.com}, Zhen-Jun Xiao$^{2}$\footnote{E-mail: xiaozhenjun@njnu.edu.cn}}
\affiliation{1. Henan Institute of Science and Technology, Xinxiang 453003, P.R.China\\
2. Department of Physics and Institute of Theoretical
Physics, Nanjing Normal University, Nanjing 210023, P.R.China   }%%

%\date{\today}
\begin{abstract}
In this paper, we investigated the process $pp \to W^{-}hj$ induced by the top-Higgs flavor-changing neutral current couplings at the LHC. We found that the cross section of $pp
\to W^{-}hj$ mainly comes from the resonant process $pp\to W^{-}t\to W^{-}hq$ due to the
anomalous $tqH$ couplings (where $q$ denotes up and charm quarks). We further studied the
observability of the top-Higgs flavor-changing neutral current couplings through the process $pp \to W^{-}(\to \ell^{-}
\bar{\nu}_{\ell}) h( \to \gamma\gamma) j$ and found that the branching ratios $Br(t\to qh)$
can be probed to $0.16\%$ at $3\sigma$ level at 14 TeV LHC with an integrated luminosity of
3000 fb$^{-1}$.
\end{abstract}


\maketitle

%%====================================================================
%\newpage
\section{Introduction}
The discovery of a 125 GeV Higgs boson at
CERN's LHC Run-I \cite{atlas,cms} has heralded the
beginning of a new era of Higgs physics. So far the observed signal
strengths, albeit with large experimental uncertainties, are found to be in good agreement with the
predictions of the Standard Model (SM) \cite{p1,p2}. Thus, precise measurements of Higgs
boson couplings to the SM particles will be a dominant task at the LHC in the next decades.

The top quark, the heaviest element of the SM, owns the largest Yukawa coupling to the Higgs
boson and has the preference to reveal the new interactions at the electroweak scale \cite{t1,t2,t3,t4}.
Because of the Glashow-Iliopoulos-Maiani mechanism \cite{gim}, the SM can only contribute
to the top-quark flavor-changing neutral current~(FCNC) at loop level with the expected
branching ratios around $10^{-15}-10^{-12}$ \cite{sm-fcnc}.
However, such suppression can be relaxed by the extended flavor structures, and consequently
larger branching ratios of $t\to qH$ or $t\to qV$ are expected in many new physics (NP) models,
  for example, in the minimal supersymmetric standard model \cite{rpc,rpv}, the two-Higgs-doublet
models \cite{2hdm-1,2hdm-2}, extra dimensions \cite{ED}, little Higgs with T-parity \cite{lht}
model, and/or the other miscellaneous models \cite{other1,other2,other3}.
So the measurement of any excess in the branching ratios for top-quark FCNC processes would be an
indication to the NP beyond the SM.

At the LHC, top quark can be copiously produced in pair
production, which leads to a high precision for the study of top observables, such as its couplings and rare
decays~\cite{tfcnc-th,multilepton,tfcnc-exp,atl-2013,atlas-fcnc,cms-fcnc,13112028,wlei-jhep,prd92-074015,160204670,prd92-113012,jhep-15-061}.
On the other hand, the single top and Higgs associated production process $pp\to th$ is also important to search for
the FCNC top-Higgs couplings, which has been emphasized in Refs.~\cite{prd86-094014,th-1,th-2,th-3,th-5,th-6}.
Up to now, the ATLAS and CMS collaborations have set the upper limits of $Br(t\to qH)<0.79\%$ \cite{atlas-fcnc} and
$Br(t\to cH) < 0.56\%$ at $95\%$ C.L. \cite{cms-fcnc} using the processes
$pp\to t\bar{t}\to Wb+qh\to \ell \nu b +q\gamma\gamma$.
The author of Ref.~\cite{prd86-094014} studied the anomalous production of $th$ via the process
$pp\to th\to (Wb) + h\to (\ell \nu) b + (b\bar{b})$ at the LHC including complete QCD next-to-leading
order (NLO) corrections, where the upper limits on the branching ratios of top-quark rare
decays are obtained:~$Br(t\to hu)\leq 4.1\times 10^{-4},~Br(t\to hc)\leq 1.5\times 10^{-3}$.

Compared with the top-quark pair production at the LHC, single top production is a very important process because its cross sections are directly
proportional to the top's weak couplings. At the LHC, the top quark can
be singly produced associated with $W$ bosons ($tW$) via the process $bg\to tW^{-}$.
The cross section is predicted at the NLO plus the contribution due to the resummation of soft-gluon
bremsstrahlung at NNLL order to be about 41.8 pb at the 14 TeV LHC for $m_t=173$ GeV \cite{tw-nnll}. It has
been shown that this process is observable at the LHC using the
fully simulated data at the CMS and ATLAS detectors \cite{tw-cms,tw-atlas}. Very recently,
Ref.~\cite{151106748} has given an experimental review of the study of processes with a
single top quark at the LHC, where the $Wtb$ coupling (see, for example,
Refs.~\cite{twb1,twb2}) and other FCNC couplings \cite{f1,f2,tw-tqg} are sensitive to the single
top production processes. Although the top pair production process has the largest cross section,
other processes with different production mechanisms and final states can also be studied independently and compared with the top-quark pair production.
It is therefore worthwhile to perform a complete calculation of the $pp\to W^{-}t\to W^{-}hq$ process
due to the presence of the top-Higgs FCNC couplings $tqH$. It is a kind of the complementary processes to be
studied in addition to the top pair and $th$ production processes.

This paper is arranged as follows. In Sec. II, we briefly describe the top-Higgs FCNC interactions
and review the current limits on top-Higgs FCNC processes from direct and indirect searches. In Sec.
III, we discuss the observability of the top-Higgs FCNC couplings through the process $pp \to W^{-}hj$
at 14 TeV LHC. Finally, a short summary is given in Sec. IV.

\section{Top-Higgs FCNC couplings and current constraints}

Considering the FCNC Yukawa interactions in the framework of the effective field theory, the SM
Lagrangian can be extended simply by allowing the terms~\cite{plb703-306}
\begin{equation}
{\cal L}= \lambda_{tuh}\bar{t}Hu+\lambda_{tch}\bar{t}Hc+h.c.,
\label{tqh}
\end{equation}
where the real parameters $\lambda_{tuh}$ and $\lambda_{tch}$ denote the FCNC couplings of
$\bar{t}Hu$ and $\bar{t}Hc$. Thus, the total decay width of the top-quark $\Gamma_t$ can be written as the form of
\begin{eqnarray}
\Gamma_t = \Gamma^{SM}(t\rightarrow W^-b)+\Gamma(t\rightarrow qH),
\end{eqnarray}
where $q=u,c$. The decay width of the dominant top quark decay mode $t\rightarrow W^-b$ at the
leading order (LO) and the NLO
can be found in Ref.~\cite{twb}. After assuming the dominant top-decay width $t \to bW$  and neglecting
the up- and charm-quark masses, the branching ratio of $t \to qh$ can be approximately given by the
form \cite{th-1}:
\begin{equation}
Br(t \to qh) = \frac{\lambda^{2}_{tqh}}{\sqrt{2} m^2_t G_F}\frac{(1-x^2_h)^2}{(1-x^2_W)^2
(1+2x^2_W)}\kappa_{QCD} \simeq 0.58\lambda_{tqh}^{2},
\end{equation}
where $G_F$ is the Fermi constant, $x_W=m_W/m_t$ and $x_h=m_h/m_t$. The factor
$\kappa_{QCD}$ is the NLO QCD correction to $Br(t \to qh)$, which is calculated as
$\kappa_{QCD}=1+0.97\alpha_{s}\simeq 1.1$ by the results of high-order corrections to $t\to bW$~\cite{twb}
and $t \to ch$~\cite{nlo}.

\begin{table}[htb]
\begin{center}
\caption{The most stringent upper bounds on the $Br(t\to qH)$ at $95\%$ C.L. obtained from the ATLAS and
CMS collaborations and other phenomenological studies from different channels. \label{limit}}
\vspace{0.2cm}
\begin{tabular}{|c|c|c|}
\hline
 Limits on branching ratio & Search channel & Data set  \\ \hline
$Br(t\to qH)< 0.79\%$&ATLAS $t\bar{t}\to Wb+qH\to \ell \nu b +\gamma\gamma q$&4.7,20 fb$^{-1}$ $@$ 7,8 TeV \cite{atlas-fcnc}\\
 \hline
$Br(t\to cH)<0.56\%$& \multirow{2}{*}{CMS $t\bar{t}\to W^{+}b+qH\to \ell \nu b +\gamma\gamma q$ } & \multirow{2}{*}{ 19.5 fb$^{-1}$ $@$ 8 TeV \cite{cms-fcnc}}\\
$Br(t\to uH)< 0.45\%$&&\\ \hline
$Br(t\to qH)<5\times10^{-4}$& ATLAS $t\bar{t}\to W^{+}b+qH\to \ell \nu b +\gamma\gamma q$  & 3000 fb$^{-1}$ $@$ 14 TeV \cite{atl-2013}\\ \hline
$Br(t\to cH)< 0.19\%$& \multirow{2}{*}{LHC $th\to \ell^{+} \nu b +\tau^{+}\tau^{-}$}  &  \multirow{2}{*}{100 fb$^{-1}$ $@$ 13 TeV  \cite{th-1}}\\
$Br(t\to uH)< 0.15\%$&&\\ \hline
$Br(t\to cH)< 0.33\%$& \multirow{2}{*}{LHC $th\to \ell^{+} \nu b +\ell^{+}\ell^{-}X$}  &  \multirow{2}{*}{100 fb$^{-1}$ $@$ 13 TeV  \cite{th-1}}\\
$Br(t\to uH)< 0.22\%$&&\\ \hline
$Br(t\to cH)< 0.48\%$& \multirow{2}{*}{LHC $th\to jj b +b\bar{b}$}  &  \multirow{2}{*}{100 fb$^{-1}$ $@$ 13 TeV  \cite{th-1}}\\
$Br(t\to uH)< 0.36\%$&&\\ \hline
$Br(t\to cH)< 0.15\%$& \multirow{2}{*}{LHC $th\to \ell \nu b +b\bar{b}$ at the NLO}  &  \multirow{2}{*}{100 fb$^{-1}$ $@$ 14 TeV  \cite{prd86-094014}}\\
$Br(t\to uH)< 0.041\%$&&\\
\hline
$Br(t\to qH)< 0.12\%$& \multirow{2}{*}{LHC $t\bar{t}\to W^{+}b+qH\to \ell^{+} \nu b +\gamma\gamma q$ } & \multirow{2}{*}{ 3000 fb$^{-1}$ $@$ 14 TeV \cite{wlei-jhep}}\\
$Br(t\to cH)< 0.26\%$&&\\ \hline
$Br(t\to qH)< 0.238\%$& LHeC $e^{-}p\to \nu_{e}\bar{t}\to \nu_{e}\bar{q}H\to \nu_{e}\bar{q}+b\bar{b}$  &  100 fb$^{-1}$ $@$ 150 GeV for $e^{-}$ \cite{160204670}\\ \hline
$Br(t\to qH)< 0.112\%$& ILC $e^{+}e^{-}\to t\bar{t}\to tqH\to \ell^{+} \nu b +b\bar{b} q$  &  3000 fb$^{-1}$ $@$ 500 GeV \cite{prd92-113012}\\ \hline
\end{tabular} \end{center}\end{table}

In Table~\ref{limit}, we summarize the most stringent upper bounds on the top-Higgs FCNC branching ratios
at $95\%$ C.L. obtained from the ATLAS and CMS collaborations and other phenomenological studies from different channels.
 From Table ~\ref{limit}, one can see that the stringent constraints of $Br(t\to cH)<0.56\%$ and $Br(t\to uH)< 0.45\%$
were reported  by the CMS Collaboration obtained by a combination of the
multilepton channel and the diphoton plus lepton channel \cite{cms-fcnc}. Furthermore, the upper limits on the flavour-changing top-quark decays can be significantly improved with the high luminosity. For example, Ref \cite{atl-2013} quotes a $95\%$ C.L. limit sensitivity in the $tt\to Wb+hq\to \ell \nu b+\gamma \gamma q$
final state of $Br(t\to hq)<5\times 10^{-4}$ with an integrated luminosity of 3000 fb$^{-1}$ at $\sqrt{s}=14$ TeV.
On the other hand, the low-energy observables, such as $D^{0}-\bar{D^{0}}$ mixing~\cite{prd81-077701} and
$Z\to c\bar{c}$~\cite{prd72-057504} can also be used to constrain
the top-quark flavor violation in the $tqH$ vertex. With the 125 GeV Higgs boson mass, upper
limits of $Br(t\to cH) < 2.1 \times 10^{-3}$ have been obtained for the $Z\to c\bar{c}$
decay~\cite{prd92-113012}. Reference~\cite{jhep06-033} derived model-independent constraints on the $tcH$ and $tuH$ couplings
that arise from the bounds on hadronic electric dipole moments.

\section{Numerical calculations and discussions}

\begin{figure}[htb]
\begin{center}
\vspace{0.5cm}
\centerline{\epsfxsize=12cm \epsffile{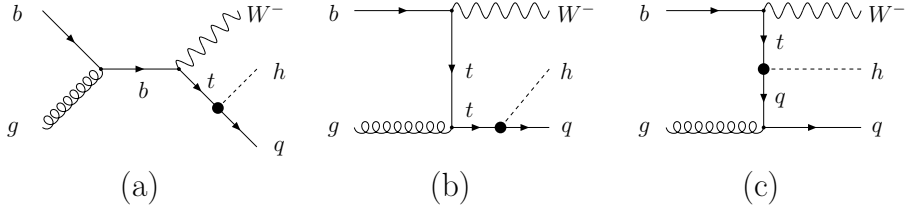}}
\caption{Feynman diagrams for the partonic process $bg \to W^{-}hq$ at the
LHC through flavor-violating top-Higgs interactions. Here $q=u,c$.}
\label{bg}
\end{center}
\end{figure}

\begin{figure}[htb]
\begin{center}
\vspace{0.5cm}
\centerline{\epsfxsize=12cm \epsffile{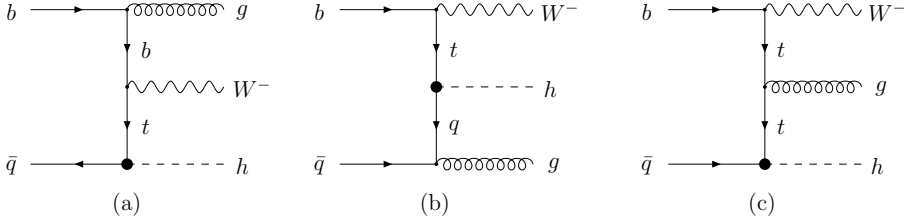}}
\caption{Feynman diagrams for the partonic process $\bar{q}b \to W^{-}hg$ at the LHC
through flavor-violating top-Higgs interactions. Here $q=u,c$.}
\label{qb}
\end{center}
\end{figure}

In this section, we search for top-Higgs FCNC couplings through the $pp\to W^{-}(\to \ell^{-}\bar{\nu})h(\to \gamma\gamma)j$
channel at the LHC, where $\ell =e, \mu$ and $q=u,c$. The Feynman diagrams are plotted in Fig.s~\ref{bg} and \ref{qb}
for the partonic processes $bg \to W^{-}hq$ and $\bar{q}b \to W^{-}hg$, respectively. In the SM, the $Wh$ + jet production
can be seen part of the inclusive $Wh$ production involving the $WWh$ coupling and the Feynman diagrams and the related
LO and NLO QCD calculations can be found, for example, in Ref.~\cite{nlo-whj}.
%%% Fig.4 %%%%%%%%%%%%%%%%%%%%
%%%%%%%%%%%%%%%%%%%%%%%%%%%%%%%%%%%%%%%%%%%%%%%%%%%%
%%%%%%%%%%%%%%%%%%%%%%%%%
%%% Fig.4 %%%%%%%%%%%%%%%%%%%%
%%%%%%%%%%%%%%%%%%%%%%%%%

There are mainly two new kinds of processes that can contribute to the production of $W^{-}hj$
at the LHC. The first one is the $bg$ fusion process $bg \to W^{-}hq$ and it is the dominant contribution,
as shown in Fig.\ref{bg}, where $hq$ can be produced not only from an on-shell top quark but also from an
off-shell top quark via the new flavor-changing couplings $tqh$. The second one is the $\bar{q}b$ fusion
process $\bar{q}b \to W^{-}hg$, as shown in Fig.~2, which is the subleading contribution.
However, such a process is affected by the initial parton sistributions dunctions(PDFs).
So, one can use this feature to disentangle the FCNC couplings of the top quark with light
quarks. Thus, it is worthwhile to perform a complete calculation of $pp \to W^{-}hj$ in the
presence of the top-Higgs FCNC couplings and explore its sensitivity to probe the top-Higgs
FCNC couplings at the LHC experiment.

We first implement the $tqH$ FCNC interactions by using the \texttt{FeynRules} package \cite{feynrules}
and calculate the LO cross section with \texttt{MadGraph5-aMC$@$NLO} \cite{mg5}. We use CTEQ6L as the
parton distribution function~\cite{cteq} and set the renormalization scale $\mu_R$ and factorization
scale $\mu_F$ to be $\mu_R=\mu_F=(m_W + m_h)/2$. In this work, we assume $\lambda_{tqh}\leq0.1$ to
satisfy the direct constraints from the ATLAS and CMS results~\cite{atlas-fcnc,cms-fcnc}. The SM
input parameters are taken as follows \cite{pdg}:
\begin{align}
m_H&=125{\rm ~GeV}, \quad m_t=173.21{\rm ~GeV}, \quad m_W=80.385{\rm ~GeV},\\ \nonumber
\alpha(m_Z)&=1/127.9, \quad \alpha_{s}(m_Z)=0.1185, \quad G_F=1.166370\times 10^{-5}\ {\rm GeV^{-2}}.
\end{align}

%%Fig.1 %%%%%%%%%%%%%%%%%%%%
\begin{figure}[thb]
\begin{center}
\vspace{-0.5cm}
\centerline{\epsfxsize=8cm \epsffile{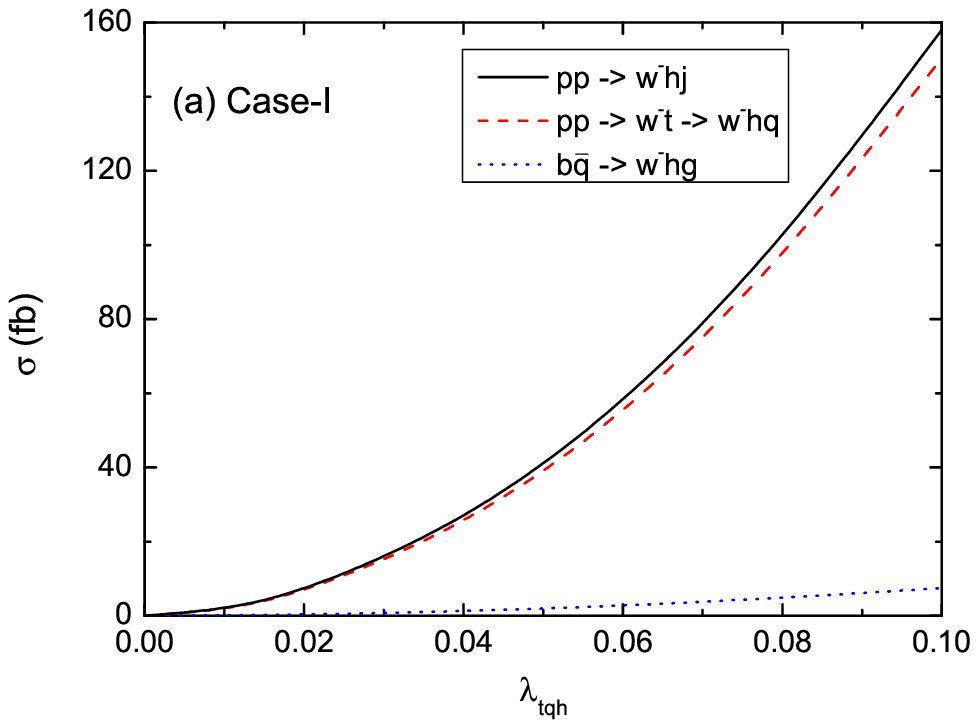}\epsfxsize=8cm \epsffile{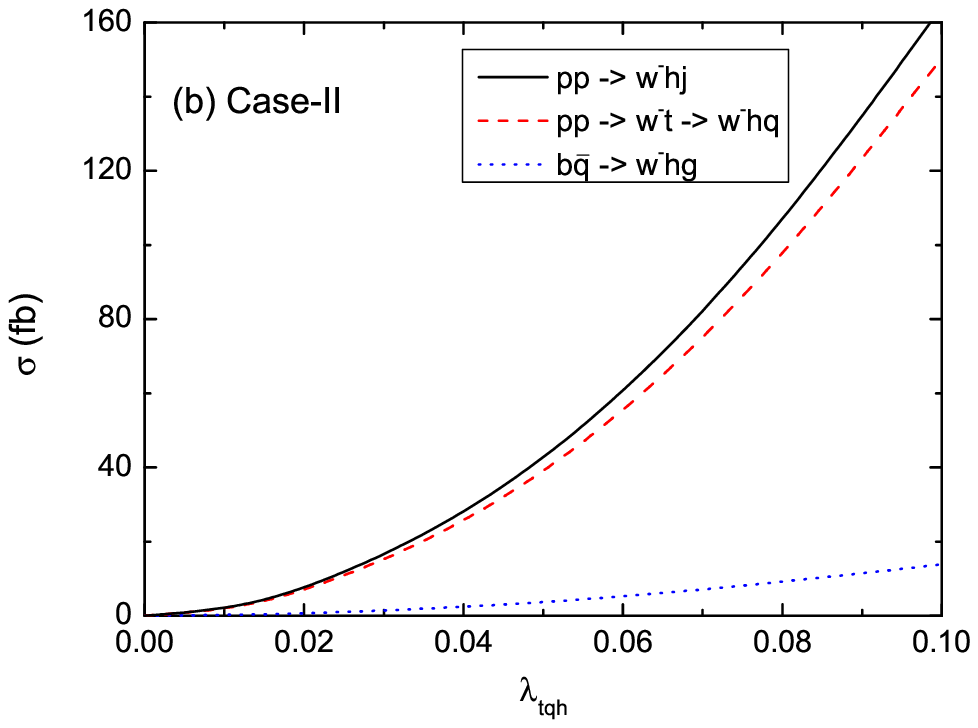}}
\centerline{\epsfxsize=8cm \epsffile{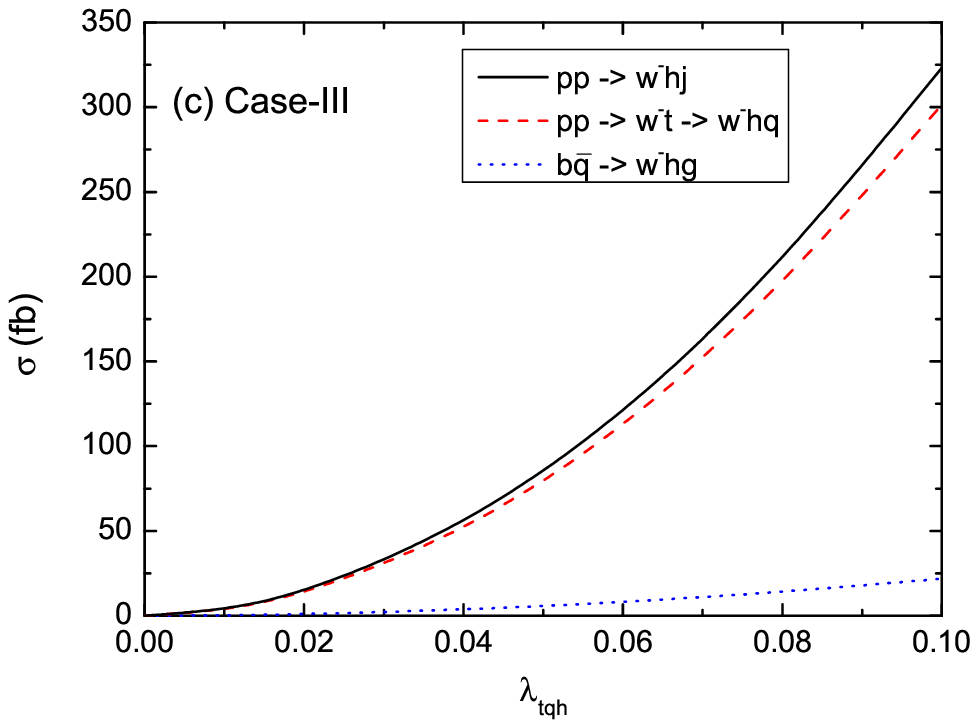}}
\caption{The dependence of the cross sections $\sigma$ at 14 TeV LHC on the top-Higgs FCNC
couplings $\lambda_{tqh}$ for Case I to Case III.}
\label{cross}
\end{center}
\end{figure}
%%%%%%%%%%%%%%%%%%%%%%%%%

In Fig.\ref{cross}, we show the dependence of the cross sections $\sigma_{Whj}$ on the
top-Higgs FCNC couplings $\lambda_{tqh}$ at 14 TeV LHC for three different cases: (I)~
$\lambda_{tqh}=\lambda_{tch}, \lambda_{tuh}=0$, (II)~$\lambda_{tqh}=\lambda_{tuh},
\lambda_{tch}=0$ and (III)~$\lambda_{tqh}=\lambda_{tuh}=\lambda_{tch}$. Since
the analysis does not distinguish between the $t\to ch$ and $t\to uh$ final states which have similar
acceptances, we take $\lambda_{tuh}=\lambda_{tch}$ in Case III for the simplicity of calculation.
From Fig.\ref{cross}, one can see that:
\begin{enumerate}[(1)]
\item
For three cases, the dominant contributions to $pp \to W^{-}hj$ are come from the $bg$-fusion
process. To be specific, when $\sqrt{s}= 14$ TeV and $\lambda_{tqh}=0.1$, the full cross
sections of $pp \to W^{-}hj$ are about 1.05, 1.10 and 1.07 times larger than that of
$pp \to W^{-}t \to W^{-}hq$ in Cases I, II and III, respectively.

\item
For the same values of $\lambda_{tuh}$ and $\lambda_{tch}$, the cross section of $pp \to W^{-}hj$
in Case I is smaller than that in Case II, since the charm-quark has the a smaller PDF than the up-quark in the proton.
Thus, for a given collider energy and luminosity, we can expect the sensitivity to the coupling
$\lambda_{tuh}$ to be better than that to $\lambda_{tch}$.

\item
For $\lambda_{tqh}=0.1$ in Cases I-III, the total cross sections of $pp \to W^{-}hj$ are about
158, 162, and 325 fb, respectively, while the cross section for the SM process $pp\to W^{-}hj$
is about 225 fb with the same input parameters and cuts. Thus, in the following calculations
we will mainly consider the probed sensitivity for the Case III due to the relative large cross section.
\end{enumerate}

Although the branching ratio is
small, the diphoton decay mode $h\to \gamma \gamma$ of the Higgs boson is one of the two cleanest channels of the Higgs
boson, which allows a sharp reconstructed peak right at the Higgs boson mass, and has the great
advantage that most QCD backgrounds are gone. Thus, in the following calculations, we perform the
Monte Carlo simulation and explore the sensitivity of 14 TeV LHC to the top-Higgs FCNC couplings $\lambda_{tqh}$
through the channel,
\begin{equation}\label{signal}
pp \to W^{-}(\to \ell^{-} \bar{\nu}_{\ell}) h( \to \gamma\gamma) j.
\end{equation}
As can be seen, in this case, the studied topology of our signal gives rise to the
jet + $\slashed E_{T}$ + diphoton signature characterized by one jet, one lepton, a missing
$\slashed E_{T}$ from the undetected neutrino and a diphoton signal appearing as a narrow
resonance centered around the Higgs boson mass.
The main SM backgrounds which yield the identical final states to the signal include two parts: the
resonant and the nonresonant backgrounds. For the former, they mainly come from the processes that
have a Higgs boson decaying to diphoton in the final states, such as $W^{-}hj$, $Zhj$ and $W^{+}W^{-}h$
productions ($j$ denotes non-bottom-quark jets), with one daughter lepton of $Z$ and one of the jets of
$W^{+}$ being mistagged. For the latter, the main background processes contain the diphoton events
associatedly  produced, such as $W^{-}j\gamma\gamma$, $Zj\gamma\gamma$ and $\bar{t}j\gamma\gamma$, where
$W^{-}j\gamma\gamma$ is the most dominant background.

All of these signal and backgrounds events are generated at leading order using
\texttt{MadGraph5 -aMC$@$NLO}. \texttt{PYTHIA} \cite{pythia} is utilized for parton
shower and hadronization. \texttt{Delphes} \cite{delphes} is then employed to account for the detector
simulations, and \texttt{MadAnalysis5} \cite{ma5} is used for analysis, where the (mis)tagging efficiencies
and fake rates are assumed to be their default values.  When generating the parton level events,
we assume $\mu_R=\mu_F$ to be the default event-by-event value. The anti-$k_{t}$ algorithm
\cite{antikt} with the jet radius of 0.4 is used to reconstruct jets. The QCD corrections for the
dominant backgrounds are considered by including a $k$ factor, which is 1.12 for $W^{-}hj$
\cite{nlo-whj}, 1.2 for $Zhj$ \cite{nlo-zhj}, and 1.3 for $W^{-}j\gamma\gamma$ \cite{nlo-wjaa,14050301}.
On the other hand, the MLM scheme~\cite{MLM} is used to match our matrix element with a parton shower.
Here, it should be mentioned that the $k$ factor for the LO cross section of $\sigma_{tW^{-}}$ is chosen
as 1.53 for the prediction at NLO+NNLL QCD corrections to $tW^{-}$ production at the 14 TeV LHC
\cite{zhu,tw-nnll}. Because of the small contribution from the process $b\bar{q}\to W^{-}hg$, it is
safe to take the LO cross section for this subleading process.

In our simulation, we generate 400,000 events for the signals and 400,000 events for the backgrounds. We first
employ some basic cuts for the selection of events:
\beq
p_{T}^{j} > 25 \rm ~GeV, \quad    p_{T}^{\ell} > 20 \rm ~GeV,\quad |\eta_{j}|< 2.5, \quad  |\eta_{\ell}|<2.0,
\quad  \Delta R_{ij}>0.4,
\eeq
where $p_{T}^{j,\ell}$ and $|\eta_{j,\ell}|$ are the transverse momentum and the pseudorapidity of the jet and leptons, respectively.
  $\Delta R=\sqrt{(\Delta\phi)^{2}+(\Delta\eta)^{2}}$ is the particle separation among the objects (the jet,
the lepton and the photons) in the final state with $\Delta\phi$ and $\Delta\eta$ being the separation in the
azimuth angle and rapidity, respectively.

  %%%%%%%%%%%%%%%%%%%%%%%%%%%%%%%%%%%%%%%%%%%%%%%%%%
\begin{figure}[htb]
\begin{center}
\vspace{-0.5cm}
\centerline{\epsfxsize=8cm\epsffile{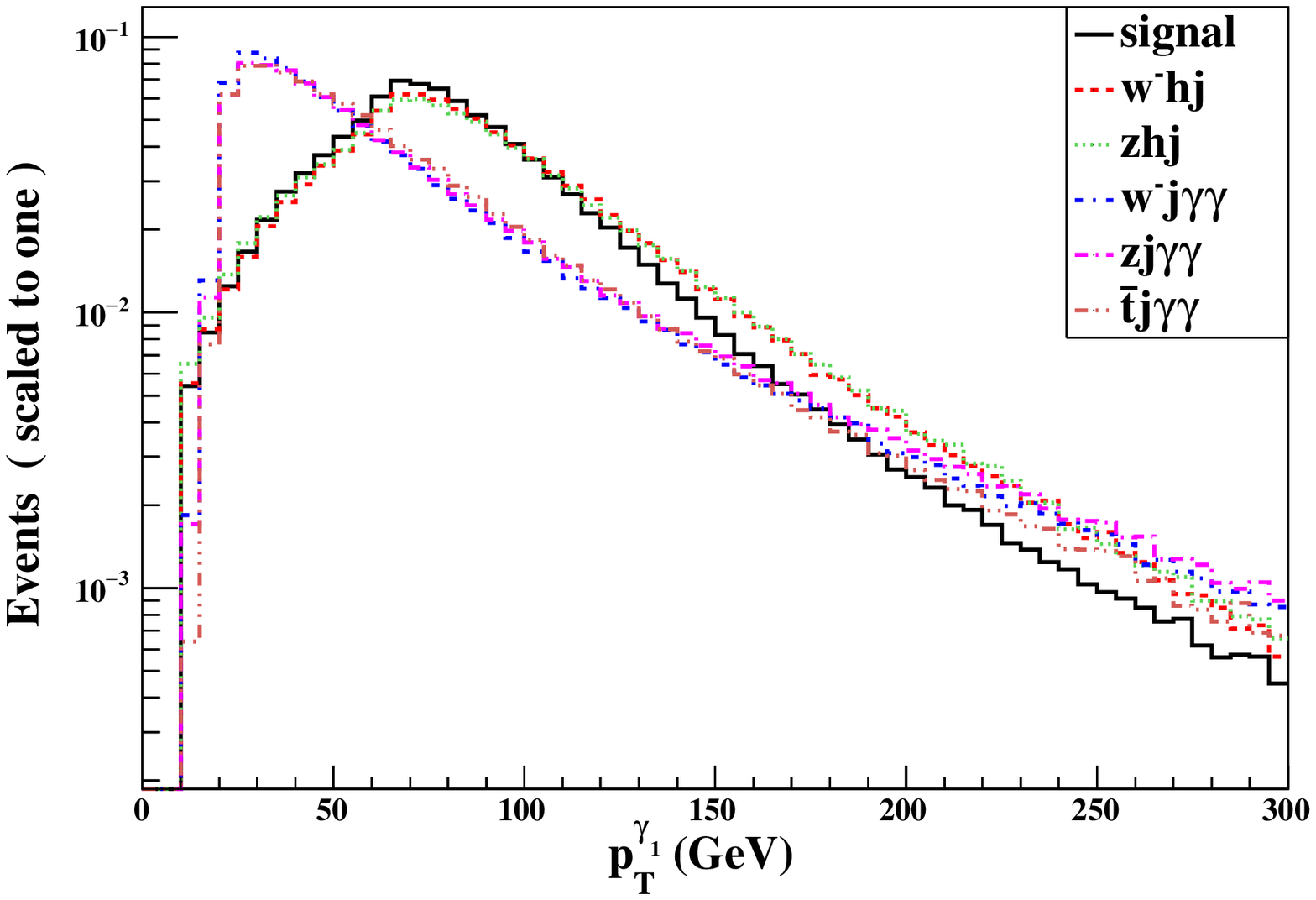}
\hspace{-0.5cm}\epsfxsize=8cm\epsffile{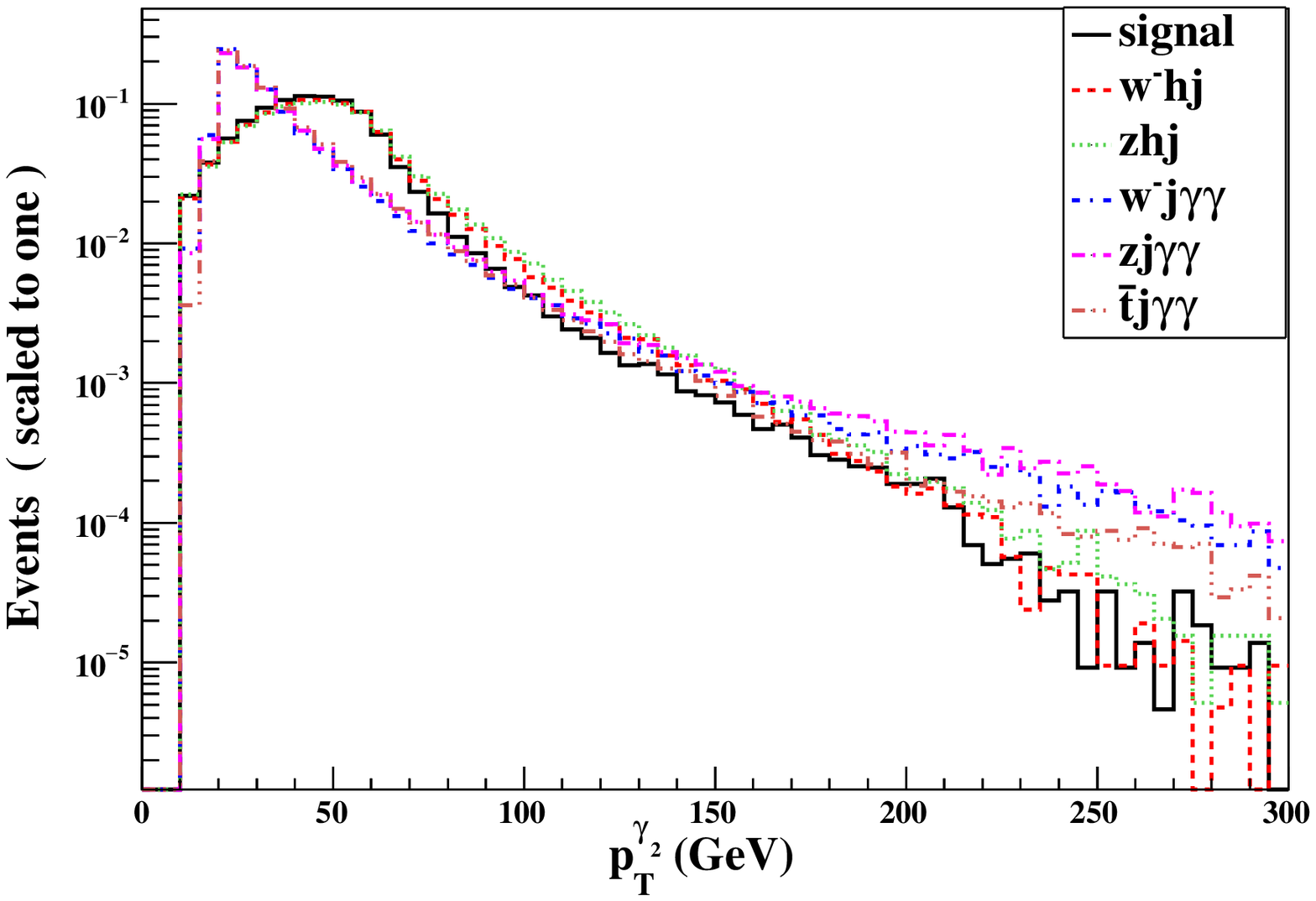}}
\centerline{\epsfxsize=8cm\epsffile{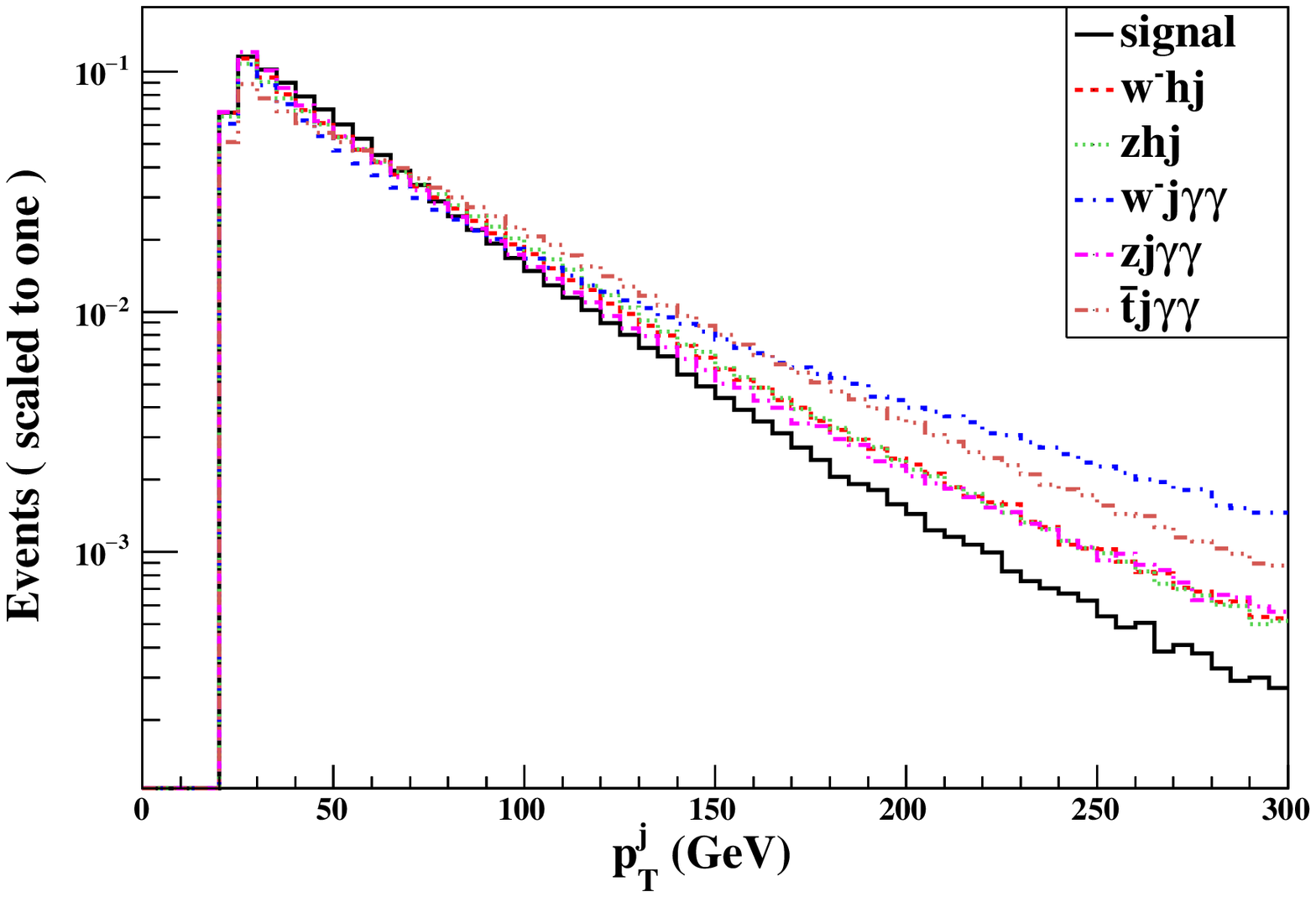}
\hspace{-0.5cm}\epsfxsize=8cm\epsffile{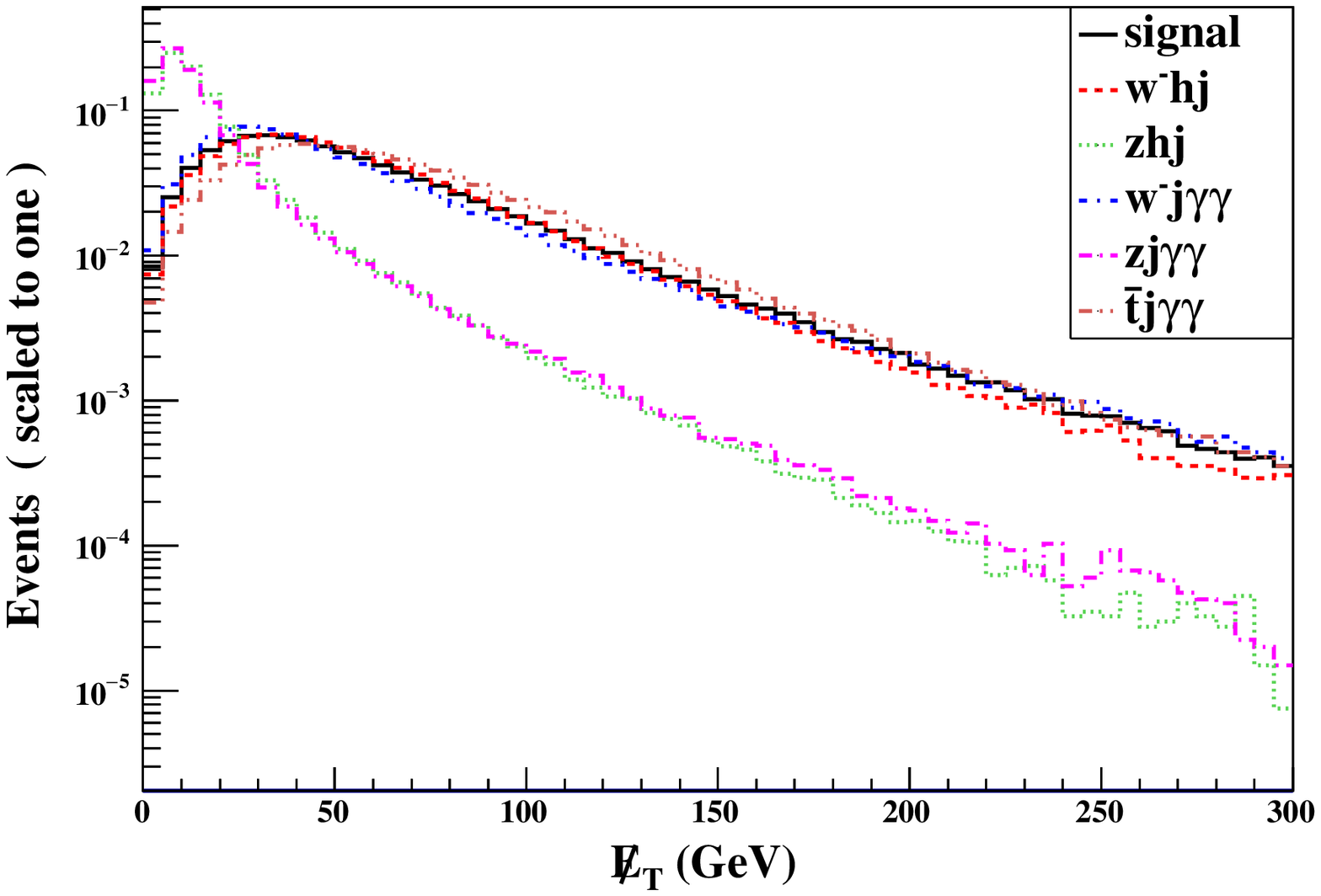}}
\vspace{0.2cm}
\centerline{\epsfxsize=8cm\epsffile{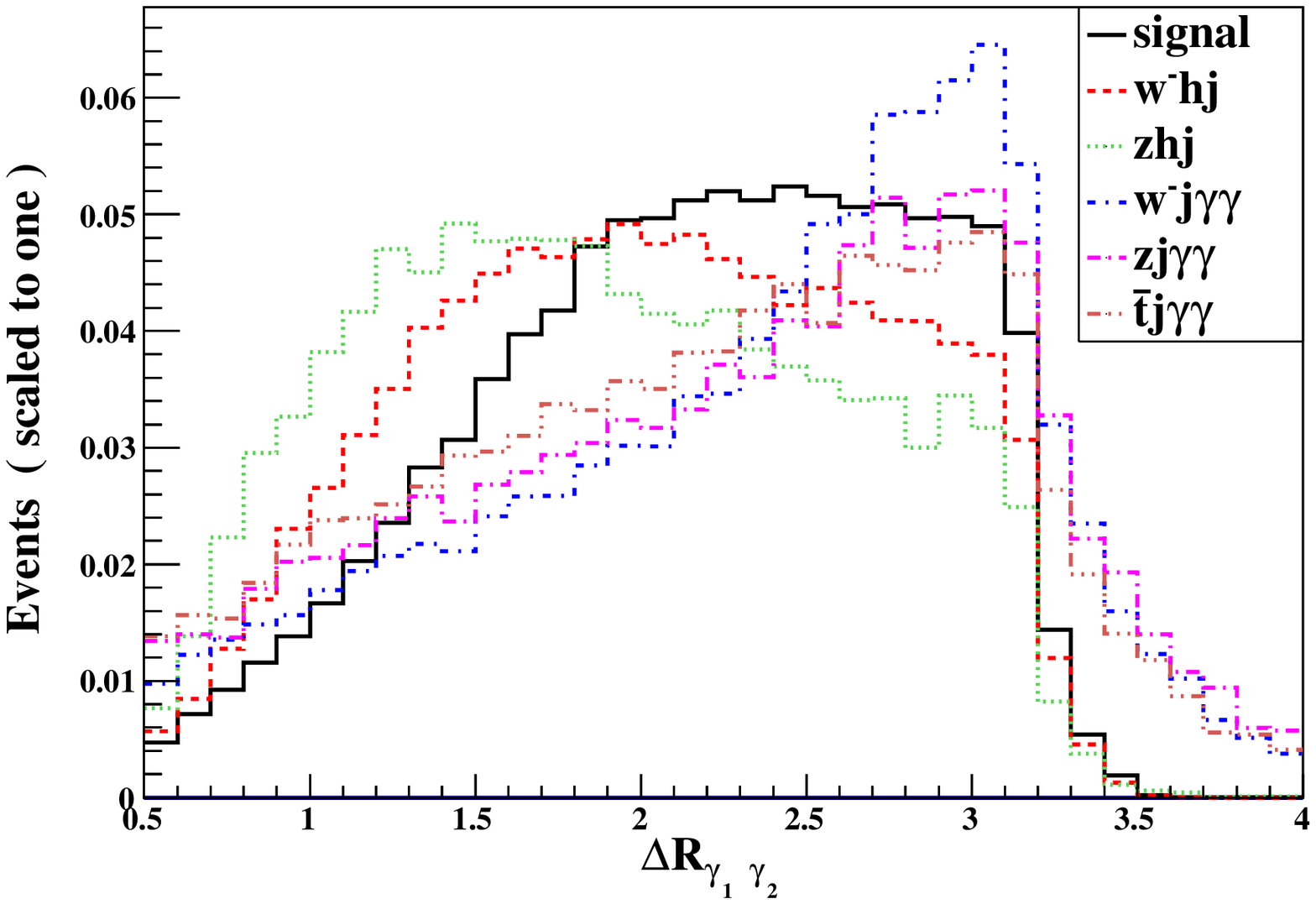}
\hspace{-0.5cm}\epsfxsize=8cm\epsffile{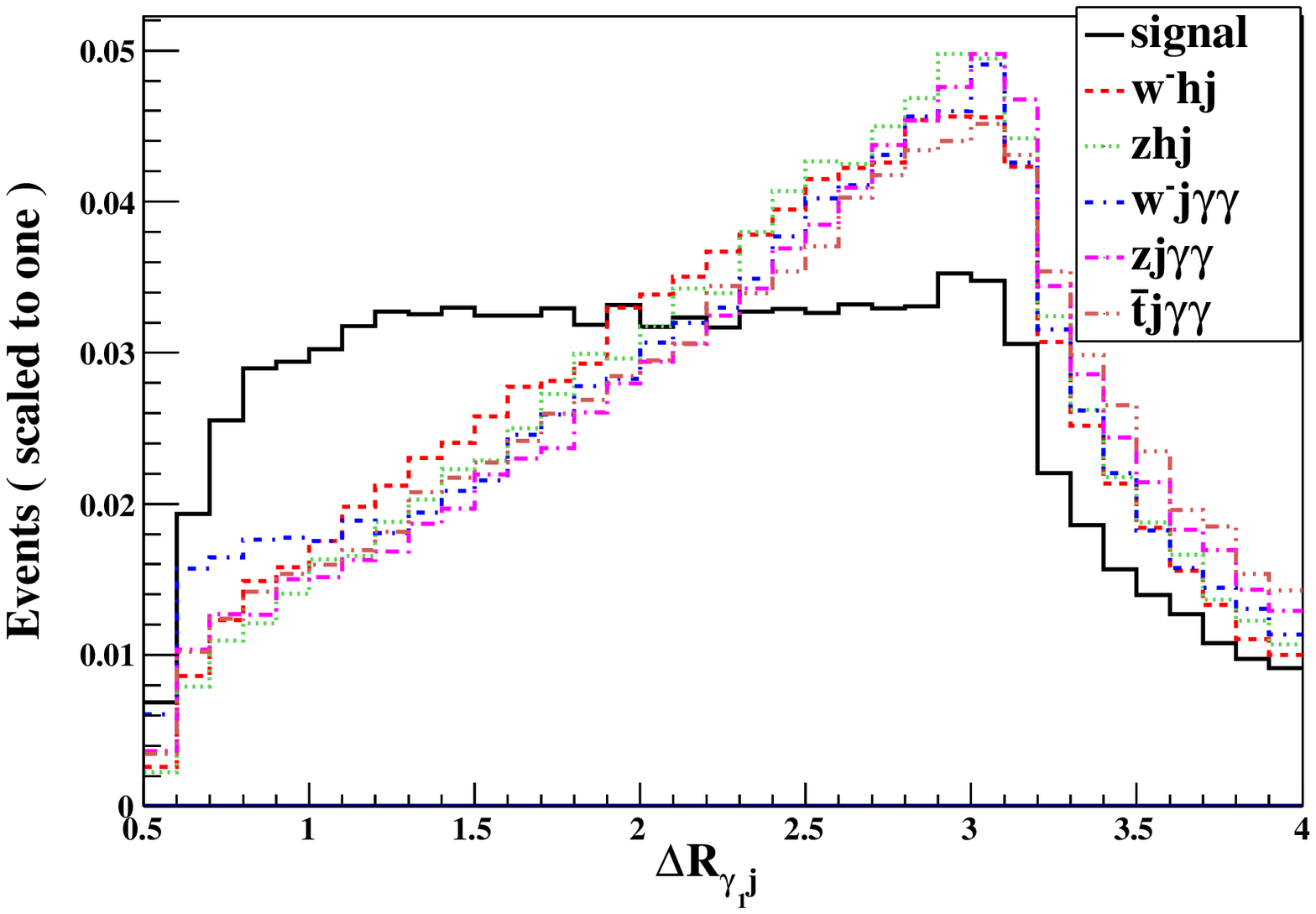}}
\caption{Normalized distributions of transverse momentum  $p_{T}^{\gamma_{1}}$, $p_{T}^{\gamma_{2}}$, $p_{T}^{j}$;
the transverse missing energy $\slashed E_{T}$; and various separations $\Delta R_{\gamma_{1}\gamma_{2},\gamma_{1}j}$
in the signals and backgrounds at 14 TeV LHC.}
\label{photon}
\end{center}
\end{figure}

%%%%%%%%%%%%%%%%%%%%%%%%%
In Fig.~\ref{photon}, we show the distributions of some kinematical variables for the signal (with the fixed parameter $\lambda_{tqh}=0.1$ in Case III) and
backgrounds at 14 TeV LHC.  The result of $W^{+}W^{-}h$ is not shown since it is subdominant with a very small cross
section (at the level of $10^{-3}$ fb) for the same final states. According to the distribution differences between
the signal and backgrounds, we can improve the ratio of signal to backgrounds by making suitable kinematical cuts.
First, since the two photons in the signal and the resonant backgrounds come from the Higgs boson, they have
peaks around $m_h/2$ and possess the harder $p_T$ spectrum than those in the nonresonant backgrounds.
The distributions of the transverse momentum $p_{T}^{j}$ of the jet share a similar feature. The missing transverse
energy $\slashed E_T$ in the backgrounds $Zhj$ and $Zh\gamma\gamma$ has
a peak at bout 10 GeV, while others share a similar feature, and thus the suitable cut on $\slashed E_T$ can largely
suppress these two backgrounds.
Thus, we can apply the following cuts to suppress the backgrounds:
\beq
p_{T}^{\gamma_1}>55 \rm ~GeV, \quad p^{\gamma_2}_{T}>35 \rm ~GeV,\quad  p_{T}^{j}<80 \rm ~GeV,\quad  \slashed E_T>20 \rm ~GeV.
\eeq
Second, since two photons in the signals are from the Higgs boson decay, the signal rates peak at their
invariant mass around the Higgs mass with relative small separation. The same features are shared by the
separation $\Delta R_{\gamma_{1}j}$ for the signal and the background, since the jet and the photon in
the signal come from the same particle (top quark). These differences between the signal
and backgrounds suggest the following comprehensive cuts:
\beq
1<\Delta R_{\gamma_{1}\gamma_{2}}< 2.7, \quad    0.5<\Delta R_{\gamma_{1}j}< 2.2.
\eeq

%%Fig. %%%%%%%%%%%%%%%%%%%%
\begin{figure}[htb]
\begin{center}
\vspace{-0.5cm}
\centerline{\epsfxsize=8cm \epsffile{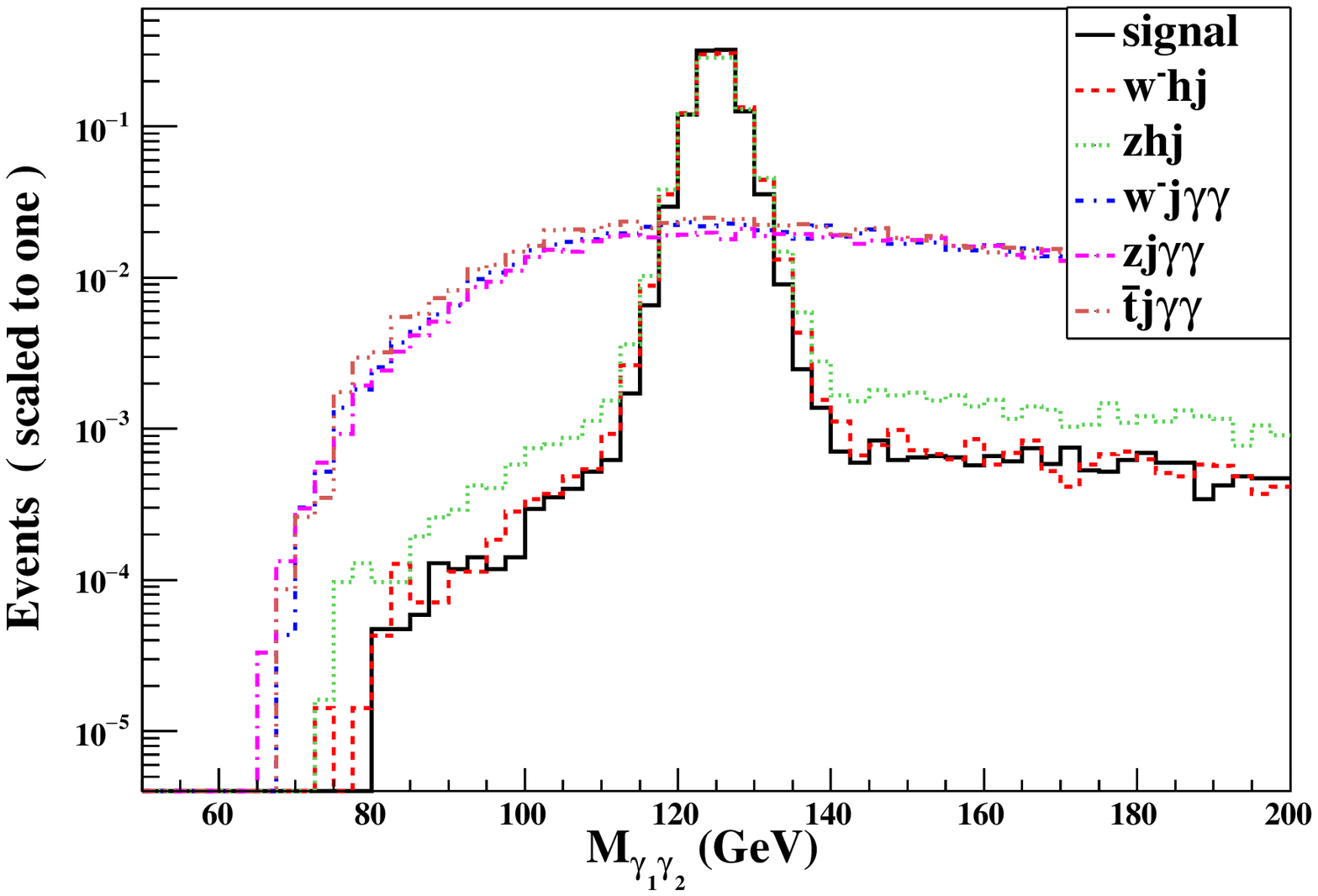}
\epsfxsize=8cm \epsffile{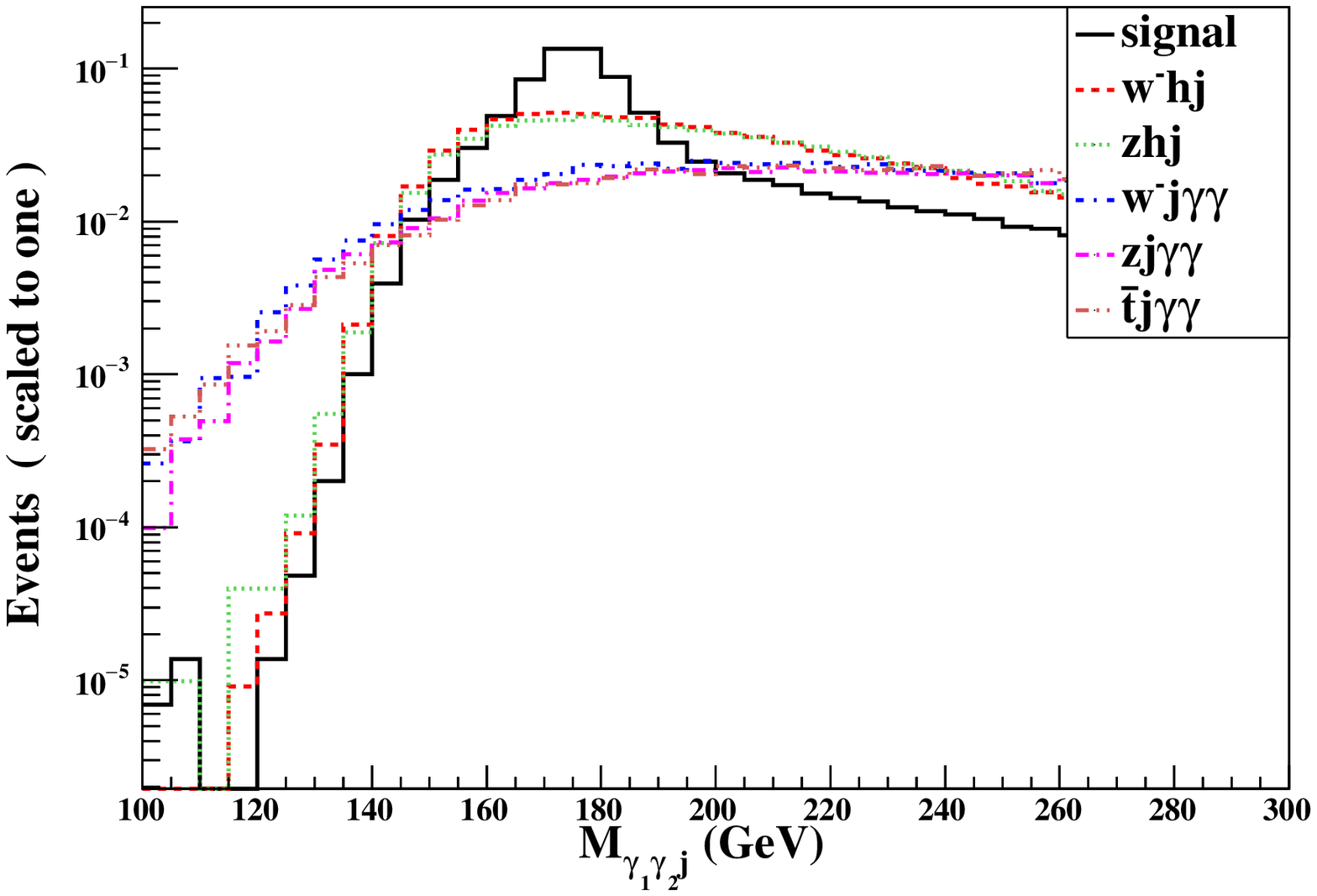}}
\caption{Normalized invariant-mass distribution of two photons (left) and the diphoton and light jet
system (right) at the 14 TeV LHC.}
\label{higgs}
\end{center}
\end{figure}
%%%%%%%%%%%%%%%%%%%%%%%%%
Next, we consider utilizing the invariant-mass distributions to further suppress the background.
Figure~\ref{higgs} shows the normalized invariant-mass distribution $M_{\gamma\gamma}$ of two photons
and $M_{\gamma\gamma j}$ of the diphoton and light jet system, respectively. Although the diphoton
decay channel has a small branching ratio, it has the advantage of good resolution on the $\gamma\gamma$
resonance and is also free from the large QCD backgrounds. Thus, we can use a narrow invariant-mass
window $|M_{\gamma\gamma}-m_H|<5$ GeV to further reduce the nonresonant backgrounds. Very similar
to $M_{\gamma\gamma}$, the invariant mass distribution of the diphoton and the light jet
$M_{\gamma_1\gamma_2 j}$ also has a peak around the top-quark mass in the signal other than
the backgrounds, which can be used to further remove the backgrounds. We therefore impose a
cut on $M_{\gamma\gamma j}$:
 \beq
|M_{\gamma\gamma j}-m_t|<10 \rm ~GeV.
\eeq

For a short summary, we list all the cut-based selections here:
\begin{itemize}
\item[(1)] Basic cut: $p_T^{j}>25$ GeV, $p_T^{\ell}>20$ GeV, $|\eta_{j}|<2.5$, $|\eta_{\ell}|<2$
and $\Delta R_{ij}> 0.4~(i,j=\ell, j, \gamma)$;

\item[(2)] Cut 1 means the basic cuts plus missing $\slashed E_T>20$ GeV, $p_T^{\gamma_{1}}>55
\rm GeV$, $p_T^{\gamma_{2}}>35 \rm GeV$ and $p_T^{j}< 80 \rm GeV$;

\item[(3)] Cut 2 means Cut 1 plus $1<\Delta R_{\gamma_{1}\gamma_{2}}< 2.7$ and $0.5<\Delta R_{\gamma_{1}\gamma_{2}}< 2.2$.

\item[(4)] Cut 3 means Cut 2 plus requiring the invariant mass of the diphoton pair to be in the range $m_h\pm5$ GeV;

\item[(5)] Cut 4 means Cut 3 plus requiring the invariant mass of the diphoton and light jet
 system  to lie in the range $m_t\pm10$ GeV.
\end{itemize}

\begin{table}[htb]
\begin{center}
\caption{The cut flow of the cross sections (in $10^{-3}$ fb) for the signal ( $W^{-}hq$ ) in Case III
and backgrounds ( $W^{-}hj$, $Zhj$, $W^{-}j\gamma\gamma$,$Zj\gamma\gamma$ and $\bar{t}j\gamma\gamma$ )
at the 14 TeV LHC. As a comparison, the corresponding results of the
resonant production $pp \to W^{-}t \to W^{-}hq$ are also listed in the brackets. \label{cutflow}}
\vspace{0.2cm}
\begin{tabular}{|c|c||c|c|c|c|c||c|}
\hline
Cuts & Signal&$W^{-}hj$ & $Zhj$ & $W^{-}j\gamma\gamma$ &$Zj\gamma\gamma$ &$\bar{t}j\gamma\gamma$ &$S/B$  \\ \hline
Basic cuts &103~(96.1)&31.7&18.6&4268&1943&195&0.016~(0.015) \\ \hline
Cut 1&25.8~(24.16)&6.3&5.6&256&33&17.5&0.081~(0.076)\\ \hline
Cut 2 &15.1~(13.99)&2.1&0.19&54.2&7.4&6.3&0.21~(0.19)\\ \hline
Cut 3& 13.23~(12.46)&1.87&0.15&3.8&0.52&0.5&1.93~(1.82)\\ \hline
Cut 4 &11.36~(10.71)&0.63&0.066&1.36&0.2&0.24&4.55~(4.29)\\
\hline
\end{tabular} \end{center}\end{table}

The cross sections of the signal and backgrounds after imposing
the cuts are summarized in Table~\ref{cutflow}.
For the numbers of the cross sections
as listed in Table ~\ref{cutflow}, the QCD corrections to the signal and backgrounds are
included by a multiplicative $k$ factor of 1.53, 1.12, 1.2, and 1.3 to the leading-order
cross sections of $tW^{-}$~\cite{zhu,tw-nnll}, $W^{-}hj$~\cite{nlo-whj},
$Zhj$~\cite{nlo-zhj}, and $W^{-}j\gamma\gamma$~\cite{nlo-wjaa,14050301}
at the 14 TeV LHC. In the last column of Table~\ref{cutflow}, we show the signal-to-background ratio $S/B$.
From the numerical results as listed in the last line of Table~\ref{cutflow},
one can see that the $W^{-}hj$ and $W^{-}j\gamma\gamma$ are two major backgrounds after applying all those
mentioned cuts. About 11 signal events can survive after the cuts
at the LHC with an integrated luminosity of 1000 fb$^{-1}$ with a signal-to-background ratio of 4.55.
We do not consider here the Cases I and II due to their relatively small production rates.

%%% Fig.5 %%%%%%%%%%%%%%%%%%%%
\begin{figure}[htb]
\begin{center}
\vspace{-0.5cm}
\centerline{\epsfxsize=11cm \epsffile{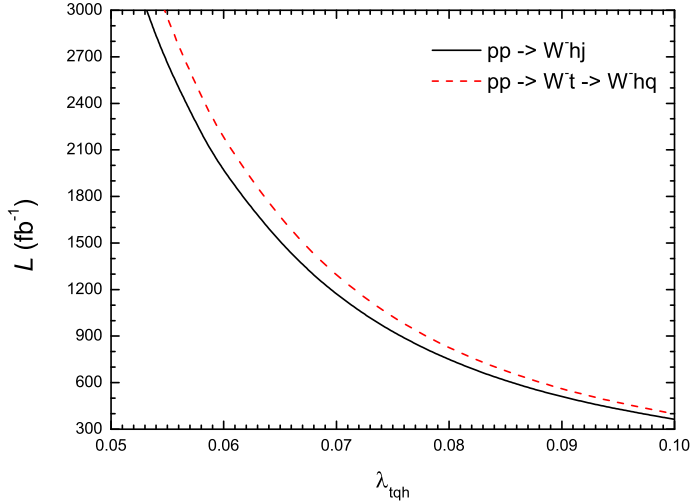}}
\caption{Contour plots in $L-\lambda_{tqh}$ plane for $3\sigma$ significance of $pp \to W^{-}hj$ (solid
line) and the resonant production $pp \to W^{-}t\to W^{-}hq$  (dashed line) at 14 TeV LHC. }
\label{ss}
\end{center}
\end{figure}

%%%%%%%%%%%%%%%%%%%%%%%%%
To estimate the observability quantitatively, we adopt the significance measurement~\cite{ss}
\beq
SS=\sqrt{2L\left [ (S+B)\ln\left(1+\frac{S}{B}\right )-S\right ]},
\eeq
where $S$ and $B$ are the signal and background cross sections and $L$ is the integrated luminosity.
In Fig.~\ref{ss}, we plot the contours of $3\sigma$ significance of $pp \to W^{-}hj$ at 14 TeV LHC
for Case III in the plane of $L-\lambda_{tqh}$. From Fig.~\ref{ss}, one can see that the flavor-changing couplings $\lambda_{tqh}$ can be probed to 0.053 at $3\sigma$ statistical sensitivity,
which corresponds to the branching ratios $Br(t\to qh)=0.16\%$ at 14 TeV LHC with $L =3000$ fb$^{-1}$.
Besides, the corresponding results of the resonant production $pp \to W^{-}t\to W^{-}hq$ are also
displayed. We can see that the LHC sensitivity to the coupling $\lambda_{tqh}$ from the full
calculation of $W^{-}hj$ production can be improved by about $6\%$ as a compared with the
resonant production $pp \to W^{-}t\to W^{-}hq$.

In comparison with the direct collider limits as presented in
Table~\ref{limit}, one can see that our results are similar to those of other studies
as given in Refs.~\cite{wlei-jhep,prd92-074015,prd92-113012,160204670}, where the sensitivity bounds
of $10^{-3}$ order are obtained through different search channels. For example,
Ref.~\cite{wlei-jhep} obtained the sensitivity bound of about $0.1-0.3\%$ through the
$pp\to t(\to \ell^{+}\nu_{\ell}b)h(\to \gamma\gamma)j$ channels at the 14 TeV with an integrated
luminosity of 3000 fb$^{-1}$.
However, the production mechanism of our search in this paper is different from the one in Ref.~\cite{wlei-jhep},
in which the authors mainly consider the top-quark pair production at the LHC via the strong interaction,
through processes such as $gg,q\bar{q}\to t\bar{t}\to \ell^{+}\nu_{\ell}b+\bar{q}H$. We expect here
to provide the complementary information for detecting the top-Higgs FCNC couplings via the
possible but different production processes at the future high-lumi LHC.
In addition, utilizing a more powerful tool, such as the multivariate technique \cite{MVA}
and various decay modes from the Higgs boson may provide better sensitivity to separate
signal from background, as shown in the very recent work \cite{160204670}.

\section{CONCLUSION}

In this paper, we investigated the process $pp \to W^{-}hj$ induced by the top-Higgs FCNC
couplings at the LHC. We found that the cross section of $pp \to W^{-}hj$ mainly comes
from the resonant process $pp\to W^{-}t\to W^{-}hq$ due to the anomalous $tqH$ couplings
(where $q$ denotes up and charm quarks). We further studied the observability of top-Higgs
FCNC couplings through the process $pp \to W^{-}(\to \ell^{-} \bar{\nu}_{\ell}) h( \to
\gamma\gamma) j$ and found that the branching ratios $Br(t\to qh)$ can be probed to
$0.16\%$ at $3\sigma$ level at 14 TeV LHC with $L =3000$ fb$^{-1}$. Compared with the
resonant production $pp \to W^{-}t\to W^{-}hq$, such a full calculation can increase
the LHC $3\sigma$ sensitivity to $Br(t \to qh)$ by about $6\%$ at 14 TeV LHC with
$L =3000$ fb$^{-1}$ because of the contribution of the nonresonant production $\bar{q}b \to W^{-}hg$.

\begin{acknowledgments}
Yao-Bei Liu thank Zhi-Long Han and Lei Wu for useful discussions about the MadGraph and
MadAnalysis5 packages. This work is supported by the National Natural Science
Foundation of China under the Grant No.11235005 and the Joint Funds
of the National Natural Science Foundation of China (U1304112).
\end{acknowledgments}

\end{document}